\documentclass[a4paper, english]{article}
\usepackage[Glenn]{fncychap}

\usepackage[utf8]{inputenc}
\usepackage{hyperref} 
\usepackage{physics}
\usepackage{graphicx}
\usepackage{fancyhdr}
\usepackage{epsfig}
\usepackage{epic}
\usepackage{bbm}
\usepackage{orcidlink}
\usepackage{slashed}
\usepackage{threeparttable}
\usepackage{amsmath} 
\usepackage{amscd}
\usepackage{amstext}
\usepackage{amssymb}
\usepackage{amsthm}
\usepackage{cancel}
\usepackage{xcolor}
\usepackage{here}
\usepackage{lscape}
\usepackage{tabularx}
\usepackage{subfigure}
\usepackage{longtable}
\usepackage{latexsym} 
\usepackage{layout}
\usepackage{float}
\usepackage{textcomp}
\usepackage{multicol} 
\usepackage{longtable} 
\usepackage{lscape} 
\usepackage{multirow} 
\usepackage{array} 
\usepackage{slashed}
\usepackage[T1]{fontenc} 
\usepackage{listings} 
\usepackage[center, small]{caption2} 
\usepackage[a4paper, total={6in, 8in}]{geometry}
\usepackage{verbatim}
\usepackage{xcite}
\usepackage{authblk}
\usepackage[square,numbers]{natbib}
\usepackage[none]{hyphenat} 
\usepackage{rotating} 
\usepackage{booktabs}

\providecommand{\keywords}[1]
{
  \small	
  \textbf{\textit{Keywords---}} #1
}

\usepackage[toc]{appendix}
\title{Geometrical BRST Quantization of Gauged Nonlinear Sigma Models: Killing Fields and Physical Cohomology}
\author[1]{
Teirungumunu Apolinar Torres Zalabata}
\author[1]{John Morales Aponte}
\author[2]{Andrés Fernando Castillo Ramírez\footnote{Corresponding author. email:ancastillo42@uan.edu.co} \orcidlink{0000-0002-9887-3293}}
\affil[1]{Departamento de F\'isica, Universidad Nacional de Colombia, Bogot\'a D.C., Colombia}
\affil[2]{Facultad de Ciencias, Universidad Antonio Nari\~no, Bogot\'a D.C., Colombia}
\date{August 2026}

\begin{document}

\maketitle

\begin{abstract}
We construct an off-shell BRST-invariant gauge-fixed formulation of a nonlinear sigma model coupled to a non-Abelian gauge field. Starting from the dynamics of GBs and their consistent couplings to gauge and ghost fields, we construct the corresponding BRST-invariant quantum theory and provide a geometrical interpretation in terms of Killing vectors. A central point of our treatment is that the Lie-bracket closure of the target-space Killing vectors provides the geometric realization of the algebra entering the BRST differential. The physical state space is characterized by the cohomology of the nilpotent BRST charge, providing a consistent separation of physical and non-physical degrees of freedom. We explicitly derive the BRST charge and present the associated symmetry transformations. Finally, we analyze the structure of the physical Hilbert space and discuss the conditions under which BRST symmetry supports unitarity and constrains the quantum effective theory. Quantum statements are conditional on a BRST-preserving measure and regulator, and the four-dimensional model is treated as an effective field theory.
\end{abstract}


\section{Introduction}
The quantum formulation of gauge theories constitutes one of the fundamental challenges in field theory, particularly for non-abelian systems endowed with local gauge symmetry due to the ambiguities arising in the selection of physical states from the Hilbert space. In this context, effective theories such as the nonlinear sigma model (NL$\sigma$M) play a central role in describing the spontaneous breaking of continuous global symmetries, the associated Goldstone bosons (GBs) and their low energy interactions, with applications to QCD and electroweak interactions~\cite{Coleman:1969sm,Callan:1969sn,peskin1995introduction,weinberg1996quantum,itzykson1980quantum}. This EFT viewpoint is standard for derivative expansions of Goldstone dynamics, including systematic loop renormalization in chiral NL$\sigma$Ms~\cite{Weinberg1979Phenomenological,GasserLeutwyler1984}.

Previous studies have formally explored the dynamics of GBs in the presence of local gauge symmetries using nonlinear realizations, covariant derivatives, and field strength tensors, establishing a geometric framework underlying chiral effective theories and gauge NL$\sigma$Ms~\cite{Longhitano1981,Henneaux:1998hq}. However, the promotion of global symmetries to local gauge symmetries introduces profound conceptual and technical challenges related to the consistent identification of physical states, unitarity, the renormalization structure of the resulting
effective theory, requiring a careful and consistent quantum formulation with a functional approach~\cite{tyutin1975gauge, kugo1979local}. The gauged nonlinear realizations can be used as electroweak chiral Lagrangians, where symmetry and power counting organize the low-energy theory after heavy scalar modes are integrated out~\cite{Longhitano1981,AppelquistBernard1980}.

The functional quantization of gauge theories via the path integral also demands the systematic elimination of the redundant gauge degrees of freedom associated with local gauge invariance~\cite{dewitt1967quantum,Nakanishi:1990qm}. A gauge-fixing procedure through appropriate conditions and the introduction of ghost fields via the FP method constitutes the starting point for constructing a well-defined quantum path integral~\cite{faddeev1967feynman, Nakanishi:1990qm}. Although FP technique is fundamental for the perturbative quantization of gauge theories, it does not by itself provide a globally unique characterization of the physical state space, as evidenced by the existence of Gribov's ambiguities and related non-perturbative constraints~\cite{singer1978gribov,gribov1978quantization}.

The BRST~(Becchi-Rouet-Stora-Tyutin) formalism emerges as a systematic framework to address the conceptual difficulties with the selection of the physical state in functional quantization. Originally introduced to preserve quantum consistency after gauge fixing, the formalism BRST defines a residual global fermionic symmetry acting on the extended field space, generated by a nilpotent charge $Q$ satisfying $Q^2=0$ and commuting with the Hamiltonian~\cite{kugo1979local,birmingham1991topological,henneaux1992quantization}. This algebraic structure allows the physical state space to be identified with the cohomology of the BRST charge $Q$, namely the space of states that are $Q$-closed but not $Q$-exact,thereby eliminating nonphysical gauge degrees of freedom from the cohomological description of
the physical sector~\cite{kugo1979local,henneaux1992quantization,becchi1976renormalization}. Under the canonical assumptions of an anomaly-free BRST symmetry, this treatment
provides the algebraic basis for gauge-independent observables and the consistent implementation of unitarity
in the physical subspace. The field-antifield formulation and algebraic renormalization provide complementary systematic frameworks for gauge fixing, BRST cohomology, admissible counterterms, and anomalies~\cite{GomisParisSamuel1995,PiguetSorella1995}.

Our work revisits the problem of constructing a BRST-consistent formulation of a NL$\sigma$M, approaching it from a geometric and cohomological perspective aimed at characterizing the physical state space through through the isometry structure generated by the Killing vectors associated with the gauged symmetries of the scalar target manifold~\cite{Henneaux:1998hq,BarnichBrandtHenneaux2000}. The framework presented here provides a unified treatment of spontaneous symmetry breaking, GBs dynamics, and the covariant quantization of gauge fields within a well-defined physical Hilbert space for the physical states. The quantum treatment of the fields in the theory is achieved by exploiting the internal geometry of the nonlinear target manifold of NL$\sigma$M \cite{CastilloRamirez:2025xey}, allowing the algebraic and geometric structures of the theory to become naturally intertwined. As a direct consequence, BRST symmetry acquires a direct geometrical interpretation, leading to a cohomological characterization of physical states \cite{PiguetSorella1995,BarnichBrandtHenneaux2000} in nonlinear gauge-invariant systems. This coordinate-covariant viewpoint is consistent with modern geometric formulations of scalar EFTs, where field-redefinition-invariant observables are expressed through geometric data of the scalar manifold~\cite{AlonsoJenkinsManohar2016}. The specific contribution is the coordinate-covariant synthesis of gauged target-space isometries, off-shell BRST symmetry, an explicit Noether charge, and the resulting physical cohomology.

To implement this framework, we systematically apply the BRST formalism to the NL$\sigma$M in the presence of a local gauge symmetry. Firstly, we consider the formal construction of the extended dynamics of the GBs in the framework of the NL$\sigma$M with the geometry underlying the gauge principle. Once the Lagrangian is constructed, we incorporate the gauge-fixing procedure, introduce the corresponding ghost and auxiliary fields, and derive the BRST transformations associated with the extended field content. Subsequently, we construct the BRST current from the extended Lagrangian and obtain the corresponding conserved BRST charge. We verify as the BRST charge nilpotency is a conequence of the algebraic structure of the Killing vectors in the target manifold. Finally, we analyze the structure of the physical state space from a cohomological perspective, emphasizing how BRST symmetry provides a consistent characterization of the physical Hilbert space and, when the symmetry is anomaly free, supports gauge-independent
observables and the quantum consistency of the EFT.

\section{Extended Dynamics of GBs}

The Lagrangian that describes the classical dynamics of GBs coupled with a local gauge symmetry is constructed from the geometric realization of spontaneously broken symmetries, coherently integrating the Goldstone modes and gauge fields through the formalism of coset spaces\footnote{To review the some results of this section, we have based this part of the text on the reference~\cite{CastilloRamirez:2025xey}}. In the following, we discuss the aspects concerning the emergence of the GBs Lagrangian in the NL$\sigma$M and its interpretation in the internal space.

\subsection{Geometrical formalism of the GBs dynamics in the internal field space}

In field theories with a continuous global symmetry group $G$ ($O(N+1)$ in the particular case of NL$\sigma$M), spontaneous symmetry breaking to a subgroup $H \subset G$ ($O(N)$) implies that the vacuum manifold is a degenerate space identified with the coset manifold $G/H$. At the level of Lie algebras, this corresponds to the decomposition $\mathcal{G}=\mathcal{H}\oplus\mathcal{I}$, where $\mathcal{H}$ contains the generators of the unbroken subgroup and $\mathcal{I}$ those associated with the broken directions. For the compact/reductive cosets considered here, $\mathcal I$ is preserved under the action of the unbroken subgroup $\mathcal{H}$. The GBs $\pi^a (x)$ are naturally associated with $\mathcal{I}$ and provide a local coordinate system on the internal manifold $G/H$. Here $G\to H$ defines the vacuum manifold $\mathcal M=G/H$, whereas only a subgroup $K\subseteq\mathrm{Isom}(\mathcal M)$ is gauged. After spontaneous symmetry breaking, the $\sigma$ field arises as a massive radial component of the order parameter. Once it is integrating out the massive radial mode of the linear sigma model, the low-energy dynamics of these fields is governed by a NL$\sigma$M whose Lagrangian takes the intrinsic geometric form

\begin{equation}
\mathcal{L}_{\text{NL}\sigma\text{M}}
=
\frac{1}{2}\, g_{ab}(\pi)\,\partial_\mu \pi^a\,\partial^\mu \pi^b,
\end{equation}

where $g_{ab}(\pi)$ is the metric induced on the coset space $G/H$ (internal space). For the symmetry-breaking pattern $O(N+1)\to O(N)$ with vacuum expectation value $v$, corresponding to $G/H\simeq S^N$, the metric reads explicitly
\begin{equation}
\label{eq:metric}
g_{ab}(\pi)
=
\delta_{ab}
+
\frac{\pi_a\pi_b}{v^2-|\pi|^2},
\end{equation}

with $|\pi|^2\equiv\delta_{ab}\pi^a\pi^b$ and $\pi_a\equiv\delta_{ab}\pi^b$.  From Eq. \eqref{eq:metric}, $g_{ab}$ encodes the intrinsic curvature of the internal space. Besides, the metric generates the characteristic nonlinear self-interactions of the Goldstone modes through the expansion around the vacuum state. Depending on the vacuum expectation value $v$, the curvature effects of the internal manifold may become significant or residual. To explicitly identify the nonlinear contributions in the NL$\sigma$M, it is necessary to expand the metric and extract the interaction terms arising from the kinetic sector of the Lagrangian. Beyond leading order, new interaction terms emerge with the expansion:
\begin{equation}
g_{ab} = \delta_{ab} + \frac{\pi_a\pi_b}{v^2-|\pi|^2}=\delta_{ab} + \frac{\pi_a\pi_b}{v^2}\left(1+\frac{|\pi|^2}{v^2}+\frac{|\pi|^4}{v^4} +\cdots\right).
\end{equation}
This chart covers $|\pi|<v$; the pole at $|\pi|=v$ is a coordinate boundary, while $S^N$ is regular with $R=N(N-1)/v^2$.

Invariance under the unbroken subgroup $H$ follows from the fact that its generators act as isometries of coset manifold $G/H$, a property that emerges by the vanishing Lie derivative $\mathcal{L}$ of the metric along the corresponding Killing vectors $\zeta_{i}$, i.e.,

\begin{equation}\label{eq:killing_inf}
\mathcal{L}_{\zeta_{i}}g_{ab} = 0
\end{equation}

If an infinitesimal symmetry transformation applies on the fields as 

\begin{equation}
\delta_\epsilon \pi^a=\epsilon^i\zeta_i^a(\pi),
\end{equation}

therefore $\zeta_i^a(\pi)$ ends up representing the infinitesimal displacement of the field along the symmetry direction labeled by $i$. From a geometrical perspective, the target-space \(K\)-orbit is generated by the gauged Killing vectors.  

However, characterizing the symmetry directions individually is not sufficient; it is also important to understand how they combine with one another. This composition is governed by the Lie bracket between Killing vectors, which encodes the Lie algebra of the acting isometry group (and, below, of the gauged subgroup $K$), not of the coset manifold $G/H$ itself:

\begin{equation}
[\zeta_i,\zeta_j]^a
=\zeta_i^b\partial_b\zeta_j^a-\zeta_j^b\partial_b\zeta_i^a
=f_{ij}{}^{k}\zeta_k^a.
\label{eq:alg_killing}
\end{equation}

This relation implies that the commutator of two Killing vectors remains tangent to the same symmetry orbit. In regions where their span has constant rank, the Killing vectors define an involutive distribution\footnote{An involutive distribution is a set of tangent directions closed under Lie brackets: if two vector fields lie in the distribution, their commutator lies in it as well~\cite{Nakahara:2003nw}}, whose integral manifolds are the local group orbits by the Frobeni\"us theorem~\cite{Lee2013}: at fixed points or singular regions, the rank may decrease, leading to a non-uniform orbit structure~\cite{Bredon1972}. Physically, successive symmetry transformations therefore remain within the gauge orbit and do not generate independent transverse directions, which must have implications in the physical choice of the states of the theory.

More generally, let $K\subseteq \mathrm{Isom}(G/H)$ denote the subgroup that is actually gauged, with generators labelled by $I,J,\ldots$ and Killing vector fields $\zeta_I=\zeta_I^a(\pi)\partial_a$. From this point onward, $a,b,\ldots$ label target-space coordinates and $I,J,K,\ldots$ label the gauged Lie algebra; lower-case $i,j$ are used only when referring specifically to the unbroken subgroup $H$. We assume a smooth isometric action with constant structure constants on the regular patch used below. Localizing the transformation parameter, $\alpha^I\to\alpha^I(x)$, introduces a gauge connection $A_\mu=A_\mu^I T_I$. The induced covariant derivative on the nonlinear target coordinates is
\begin{equation}
D_\mu\pi^a=\partial_\mu\pi^a-gA_\mu^I\zeta_I^a(\pi).
\end{equation}
For the compact gauge factor we lower gauge indices with an invariant metric $\kappa_{IJ}$ (taken as $\delta_{IJ}$ in an orthonormal basis), so $X_I=\kappa_{IJ}X^J$. The locally gauge-invariant scalar Lagrangian is therefore
\begin{equation}
\mathcal L_\pi=\frac12 g_{ab}(\pi)D_\mu\pi^aD^\mu\pi^b,
\end{equation}
where invariance follows from the Killing condition and closure written above. The $H$-specific Eq.~\eqref{eq:killing_inf} is not sufficient by itself if the gauged subgroup $K$ contains directions outside $H$. Moreover, the formal target-space measure is invariant under gauged isometries

\begin{equation}
[\mathcal D\pi]_g\equiv\prod_x\sqrt{\det g(\pi(x))}\,d^n\pi(x),\qquad \mathcal L_{\zeta_I}g=0\Rightarrow\nabla_a\zeta_I^a=0,
\end{equation}

And hence regulator-dependent anomalies remain a separate quantum issue. This qualification is essential because quantum anomalies may originate in the functional measure and are constrained by Wess-Zumino consistency conditions~\cite{PiguetSorella1995,WessZumino1971, Fujikawa1979,Brandt:1989gv}.

The field-strength tensor is the curvature of the spacetime gauge connection,
\begin{equation}
F_{\mu\nu}^{I}=\partial_\mu A_\nu^{I}-\partial_\nu A_\mu^{I}+g f_{JK}{}^{I}A_\mu^J A_\nu^K,
\end{equation}
and must be distinguished from the intrinsic Riemann curvature $R^{a}{}_{bcd}[g]$ of the target-space metric. For a compact gauge factor with the standard invariant quadratic form,
\begin{equation}
\mathcal L_{\rm YM}=-\frac14 F_{\mu\nu}^{I}F_I^{\mu\nu}.
\end{equation}
Thus the extended Goldstone and gauge dynamics can be written as
\begin{equation}\label{eq:gbosons}
\mathcal L_{GB}=\frac12 g_{ab}(\pi)D_\mu\pi^aD^\mu\pi^b-\frac14 F_{\mu\nu}^{I}F_I^{\mu\nu}.
\end{equation}
Geometrically, $A_\mu^I\zeta_I$ selects a tangent direction to the gauge orbit at each spacetime point; the BRST construction will promote these orbit directions to Grassmann-valued directions through the ghost fields.

Hence, the Lagrangian $\mathcal{L}_{GB}$ in Eq.~(\ref{eq:gbosons}) describes a theory in which the scalar fields $\pi^a$, associated with GBs, interact with the gauge fields $A_\mu^I$ through covariant derivatives and field-strength tensors. The factor $g_{ab}(\pi)$, interpreted as the metric on the internal field target space, defines the geometrical structure of the configuration space, which is typically identified with a coset manifold $G/H$, or specifically, in the NL$\sigma$M case, with the quotient space $O(N+1)/O(N).$

\section{Gauge fixing and ghost fields in quantization}~\label{sec:gfghosts}

The quantization procedure of a gauge theory requires explicitly addressing the structural redundancy arising from invariance under local gauge transformations. Although it is essential for the consistent formulation of field theories with local symmetry, this invariance introduces non-physical degrees of freedom that must be carefully eliminated to avoid ambiguities in the definition of the functional integral of the theory~\cite{peskin1995introduction, tyutin1975gauge}.

From a geometrical perspective on the target internal spaces, which is important for the subsequent development of the NL$\sigma$M, this redundancy in the field variables manifests itself through the existence of gauge orbits—equivalence classes composed of field configurations related by gauge transformations. Perturbatively, the gauge-fixing procedure selects a local representative by imposing a condition transverse to the gauge orbit. A globally unique representative need not exist because Gribov copies can obstruct the existence of a global gauge section~\cite{singer1978gribov, gribov1978quantization, henneaux1992quantization}. Accordingly, the construction developed below is local and perturbative in field space; global Gribov issues lie outside its scope.

From a general perspective, the gauge fixing is implemented by demanding a gauge condition on the gauge field $A_\mu^I(x)$, which may take different structures depending on the physical context and computational level convenience (e.g., Lorenz, axial, or Coulomb gauges). Although the gauge-fixing procedure explicitly breaks the original gauge invariance, it enables a systematic definition of the functional measure in the path integral, and constitutes an essential step toward the implementation of the Faddeev–Popov (FP) quantization procedure and its BRST extension. These formalisms restore the consistency of the quantum treatment through the introduction of ghost and auxiliary fields together with residual symmetries on fermion symmetries~\cite{tyutin1975gauge, becchi1976renormalization,PiguetSorella1995,faddeev1967feynman}.

To implement the gauge-fixing procedure, we begin with the functional expression for the transition amplitude (generating functional) of a pure YM theory described by the Lagrangian $\mathcal{L}_{YM}$, given by the path integral

\begin{equation}\label{eq:Z_inicio}
Z = \int \left[ \mathcal{D}A \right] e^{i\int d^4x\, \mathcal{L}_{YM}(A)},
\end{equation}
where $[\mathcal{D}A]$ is the functional integration measure over all possible configurations of the gauge field $A_\mu^{I}(x)$, where $\bar{a}$ is a Lie algebra index. However, the functional integral over gauge fields includes an overcounting of physically equivalent configurations. This redundancy arises because the gauge field $A_\mu^{I}(x)$ does not represent a physical observable by itself, but it is defined modulo local gauge transformations, which form a continuous group of internal symmetries. Under an infinitesimal gauge transformation and normalizing over gauge parameter, the field changes according to:

\begin{equation}
A_\mu^{\prime I}=A_\mu^I+(D_\mu\alpha)^I
=A_\mu^I+\partial_\mu\alpha^I+g f_{JK}{}^I A_\mu^J\alpha^K.
\end{equation}

where $\alpha^{I}(x)$ is the normalized gauge parameter.  The constants $f^{I_{, JK}}$ are the structure constants of the gauged Lie algebra.
The gauge symmetry implies that many configurations of the gauge field correspond to the same physical state, introducing a degeneracy in the functional integral; that is, the integration is performed multiple times over equivalent points belonging to the same gauge orbit. 

The main objective now is to eliminate the redundancy in the gauge, a gauge-fixing condition $G^{I}(A)=0$ must be imposed. This procedure allows to select a local representative near a regular gauge orbit rather than a globally unique representative. Global uniqueness is not assumed because Gribov copies may remain. This condition restricts the domain of the functional integration to a hypersurface transversal to the gauge orbits, thereby enabling a consistent quantization \cite{singer1978gribov, gribov1978quantization,faddeev1967feynman}.

To implement the method for gauge fixing, we use the FP identity~\cite{faddeev1967feynman}:

\begin{equation}\label{eq:FP_identity}
1 = \int \left[\mathcal{D}\theta \right] \delta\big(G(A^\theta)\big) \det\left(\frac{\delta G(A^\theta)}{\delta \theta}\right),
\end{equation}

where $A^{\theta}$ denotes the gauge-transformed field. This identity allows one to insert the gauge-fixing condition into the path integral without modifying its value, by properly adjusting the integration measure. By inserting this identity into the original path integral of Eq.~\eqref{eq:Z_inicio}, partition function becomes:

\begin{equation}
Z = \int [\mathcal DA]\,\delta(G^I(A))\,\det\!\left[\partial^\mu(D_\mu)^I{}_{J}\right]
 e^{i\int d^4x\,\mathcal L_{YM}(A)},
\end{equation}

where $D_\mu$ is the covariant derivative in the adjoint representation:

\begin{equation}
(D_\mu c)^I=\partial_\mu c^I+g f_{JK}{}^I A_\mu^Jc^K
\equiv (D_\mu)^I{}_{J}c^J.
\end{equation}

Here $c^{I}(x)$ is an anticommuting scalar field valued in the gauge algebra. To regularize this condition, one introduces an arbitrary function $w^I(x)$ and considers a family of gauge-fixing conditions $G^{I}(A)=\partial^\mu A_\mu^I(x) - w^{I}(x)$, which allows for an extension of the Lorenz gauge. By integrating over all possible functions $w^{I}(x)$ with a Gaussian weight $e^{-i\int d^4x \frac{w^I w_{I}}{2 \xi}}$, the functional delta is replaced by the corresponding quadratic gauge-fixing term.
The functional determinant  can be rewritten as an integral over new anticommuting scalar fields $c^{I}(x)$ and $\bar{c}^{I}(x)$, known as ghost fields. These are introduced through an identity based on Grassmann variables:

\begin{equation}
\mathcal M^I{}_{J}=\partial^\mu(D_\mu)^I{}_{J},\qquad
\det\mathcal M\propto\int[\mathcal Dc][\mathcal D\bar c]\,e^{-i\int d^4x\,\bar c_I\mathcal M^I{}_{J}c^J}.
\end{equation}

The overall sign of the FP operator changes the determinant only by a field-independent normalization; one convention is used here and below.

Faddeev-Popov ghosts are Grassmann-odd Lorentz scalars belonging to the unphysical sector of the gauge-fixed theory and do not occur as asymptotic physical states \cite{kugo1979local,henneaux1992quantization}. They arise naturally from the functional determinant and are essential for the perturbative consistency of the gauge-fixed formulation.
For the coupled NL$\sigma$M the gauge condition may depend on both $A$ and $\pi$. At the FP stage it is sufficient to define the gauge condition and FP operator,

\begin{equation}
F^I(A,\pi)=0,\qquad \mathcal M^I{}_J=\frac{\delta F^I}{\delta\alpha^J}.
\end{equation}

By incorporating all these elements into the functional integral, one obtains the expression:

\begin{equation}
Z=\mathcal N\!\int[\mathcal DA][\mathcal Dc][\mathcal D\bar c]\,
e^{\,i\int d^4x[\mathcal L_{\rm YM}-\frac{(\partial^\mu A_\mu^I)^2}{2\xi}
-\bar c_I\partial^\mu D_{\mu J}{}^I c^J]} .
\end{equation}
Here $\mathcal N$ absorbs the gauge-group volume; no independent $[\mathcal D\theta]$ remains after gauge fixing. The displayed determinant corresponds to the Lorenz choice $F^I=\partial^\mu A_\mu^I$; more general $F^I(A,\pi)$ generates Goldstone-dependent ghost vertices.
In this way, it is possible to identify the effective Lagrangian, which includes both the gauge-fixing term and the contribution from the FP ghost fields:

\begin{equation}\label{eq:L_BRST_base}
\mathcal L_{FP}=-\frac14F_{\mu\nu}^IF_I^{\mu\nu}-\frac1{2\xi}(\partial^\mu A_\mu^I)^2
-\bar c_I\,\partial^\mu(D_\mu c)^I.
\end{equation}

The Lagrangian $\mathcal{L}_{FP}$ of Eq.~\eqref{eq:L_BRST_base} forms the basis of the BRST formalism, which systematically introduces the auxiliary degrees of freedom through a global fermionic symmetry, henceforth \emph{the BRST symmetry}. Below, we will apply the gauge fixing procedure and ghost fields formalism to the extended dynamics of the GBs within NL$\sigma$M.

\section{Gauge-fixed Lagrangian for the NL$\sigma$M}\label{sec:extendedlagrangian}

As discussed previously in Section~\ref{sec:gfghosts}, path integral quantization requires gauge fixing to eliminate local redundancies, thereby introducing non-physical degrees of freedom such as the FP ghost fields. Since $\mathcal L_{FP}$ and $\mathcal L_{GB}$ both contain the Yang-Mills term, it must be included only once:

\begin{equation}\label{eq:lagrangiano BRST con bosones de goldstone}
\begin{aligned}
\mathcal L_{ext}={}&\frac12g_{ab}D^\mu\pi^aD_\mu\pi^b-\frac14F_{\mu\nu}^IF_I^{\mu\nu}-\frac{(\partial^\mu A_\mu^I)^2}{2\xi}-\bar c_I(\partial^\mu D_{\mu J}{}^I)c^J .
\end{aligned}
\end{equation}

Where gauged directions with $\zeta_I(0)\neq0$ are massive; an $R_\xi$ gauge cancels their $A_\mu$-Goldstone mixing \cite{Longhitano1981,AppelquistBernard1980, AlonsoJenkinsManohar2016}:
\begin{equation}
(M_A^2)_{IJ}=g^2g_{ab}(0)\zeta_I^a(0)\zeta_J^b(0).
\end{equation}

The Lagrangian $\mathcal{L}_{ext}$ of Eq.~\eqref{eq:lagrangiano BRST con bosones de goldstone} describes a non-Abelian gauge theory with spontaneous symmetry breaking (SSB), in which physical fields, gauge bosons and GBs, coexist with unphysical degrees of freedom, such as ghost fields. The fundamental roles of each of these terms in NL$\sigma$M are detailed as follows:

\begin{itemize}

\item The kinetic term, $\frac{1}{2}g_{ab}(\pi)D^\mu \pi^a D_\mu\pi^b$ represents the dynamics of the GBs which are excitation modes following the spontaneous breaking of a continuous global symmetry. Interaction with the gauge fields is incorporated through the covariant derivatives $D_\mu \pi^a$, ensuring invariance under local transformations. The metric $g_{ab}(\pi)$ defines the geometric structure of the configuration space; usually identified with a coset space $G/H$ that is in this case $O(N+1)/O(N)$.

\item The pure YM term, $-\frac{1}{4}F^{I}_{\mu\nu}F_{I}^{\mu\nu}$, accounts for the dynamics of the gauge fields, responsible for their propagation and self-interaction. In non-Abelian theories, the field-strength tensor $F_{\mu\nu}^I$ is the curvature of the spacetime gauge connection and includes nonlinear terms in $A_\mu^I$, reflecting the non-commutativity of the gauge-group algebra. It should not be identified with the intrinsic Riemann curvature of the target manifold.

\item The gauge-fixing term, $-(\partial^\mu A_\mu^I)^2/{(2\xi)}$ (Lorenz-type gauge), explicitly incorporates the gauge condition into the action, a fundamental requirement for consistently defining the path integral. The parameter $\xi$ is a free gauge parameter that allows different gauge choices, such as Feynman gauge ($\xi=1$) or Landau gauge ($\xi \rightarrow 0$). Its inclusion ensures that the propagator for $A_\mu^I$ field is properly defined. Near $\pi=0$, a broken direction admits $F^I=\partial\!\cdot\!A^I+\xi g\,g_{ab}(0)\zeta^{Ia}(0)\pi^b$, with the sign fixed by our $D_\mu\pi$ convention. This is the standard logic behind renormalizable $R_\xi$ gauges in spontaneously broken gauge theories; physical observables must remain gauge-parameter independent~\cite{tHooft1971,Nielsen1975}.

\item The ghost term, $-\bar c_I(\partial^\mu (D_\mu c)^{I}$, arises from the FP determinant. The fields $c^I$ and $\bar c_{I}$ are the Grassmann type and Lorentz scalars. Their contributions are required for the cancellation of gauge-dependent unphysical modes and for the BRST (and Slavnov-Taylor) identities \cite{Weinberg1979Phenomenological,
GasserLeutwyler1984,
Henneaux:1998hq,
PiguetSorella1995}. In a four-dimensional NL$\sigma$M this does not imply power-counting renormalizability; rather, BRST symmetry constrains the gauge-consistent counterterms order by order in the effective field theory expansion. As a consequence, predictions are organized for terms with $p/\Lambda_{\rm EFT}\ll1$, in a finite BRST-compatible operator basis retained at each derivative order.
\end{itemize}

Therefore, the Lagrangian $\mathcal{L}_{ext}$ of Eq.~\eqref{eq:lagrangiano BRST con bosones de goldstone} describes the non-Abelian gauge theory with GBs, incorporating the gauge fixing terms and ghost fields. Therefore, the theory Lagrangian of Eq.~\eqref{eq:lagrangiano BRST con bosones de goldstone} constitutes the foundation of the BRST formalism for the NL$\sigma$M, which is crucial for the quantum analysis of the theory, allowing in a direct way the perturbative expansion and the systematic implementation of BRST symmetry, which will be developed in the following sections.

\section{BRST Symmetry}

The BRST-extended Lagrangian of the GBs of Eq.~\eqref{eq:lagrangiano BRST con bosones de goldstone} exhibits a key feature: it is no longer invariant under the original gauge symmetry. This gauge symmetry breaking is not of dynamical origin but appears from the necessity of defining the path-integral measure in a non-redundant way. To define the path integral and eliminate the redundancies associated with gauge invariance, a gauge-fixing function is introduced, along with new unphysical degrees of freedom represented by ghost fields. Although these terms are essential for a consistent quantization, their presence provides the explicit breaking of the gauge symmetry in the original Lagrangian.

Starting from the extended Lagrangian $\mathcal{L}_{ext}$, we can unambiguously distinguish the different contributions that define the quantum form of a non-Abelian gauge theory with GBs. As previously noted in Eq.~\eqref{eq:lagrangiano BRST con bosones de goldstone}, the first two terms represents the essential physical dynamics: the first describes the interaction between the GBs and the gauge fields, while the second defines the propagation and self-interaction of the gauge fields through the field strength tensor $F_{\mu\nu}^{I}$.

However, the last two terms in Eq.~\eqref{eq:lagrangiano BRST con bosones de goldstone} reveal a fundamental based on the explicit breaking of gauge symmetry at the level of the Lagrangian, directly resulting from the gauge-fixing procedure. Specifically, the gauge-fixing term, dependent on the parameter $\xi$, imposes an additional condition on the fields $A_\mu^I$, while the term associated with the ghost fields ensures the correct compensation of the functional determinants emerging from the implementation of the FP identity.

Although the gauge-fixed action is no longer invariant under the original local gauge transformations, this apparent contradiction is resolved through the BRST formalism, which introduces a residual global nilpotent symmetry in the extended space of physical and auxiliary fields. The associated BRST charge organizes the physical cohomology. BRST cohomology is the space of BRST-closed states modulo BRST-exact states, thereby identifying states that differ only by gauge redundancy \cite{BarnichBrandtHenneaux2000}. In this aspect, unitarity needs the standard quartet or adjoint assumptions, while renormalization is controlled through BRST or Slavnov-Taylor identities \cite{tyutin1975gauge,becchi1976renormalization, PiguetSorella1995, Slavnov1972} within the EFT expansion. In what follows, we develop the systematic construction of this symmetry and its transformations.

It is possible to rewrite the gauge-fixing term by introducing an auxiliary scalar field $B^I$, without a kinetic term, in order to linearize the original quadratic term $-(\partial^\mu A_\mu^I)^2/(2\xi)$. This reformulation simplifies the analysis of symmetries and enables the covariant formulation of the BRST formalism. Let us then consider the functional equivalence:

\begin{equation}
-\frac{1}{2\xi}(\partial^\mu A^I_\mu)^2
\quad \longrightarrow \quad
\frac{\xi}{2}B_{I}B^I + B_{I} \partial^\mu A^I_\mu,
\end{equation}
where the algebraic equation of motion $B^I=-(\partial\!\cdot\!A^I)/\xi$ reproduces the original quadratic gauge-fixing term. By functionally integrating over $B^I$, the original term is recovered through the Gaussian identity:

\begin{equation}
\exp\left[-i\int d^4x\,\frac{(\partial^\mu A^I_\mu)^2}{2\xi}\right]
\propto
\int [\mathcal D B]\,
\exp\left[i\int d^4x\left(\frac{\xi}{2}B_{I}B^I+B_{I}\partial\cdot A^I\right)\right].
\end{equation}

The extended Lagrangian of Eq.~\eqref{eq:lagrangiano BRST con bosones de goldstone}, rewritten with the auxiliary field, then becomes:

\begin{equation}\label{lagrangiano_BRST_aux}
\mathcal L_{ext}=\frac12g_{ab}D^\mu\pi^aD_\mu\pi^b-\frac14F_{\mu\nu}^IF_I^{\mu\nu}+\frac{\xi}{2}B_IB^I+B_I\partial^\mu A_\mu^I-\bar c_I(\partial^\mu D_{\mu J}{}^I)c^J .
\end{equation}

This Lagrangian loses its invariance under local gauge transformations due to the explicit gauge fixing and the inclusion of ghost fields. However, it acquires a new global symmetry, the BRST symmetry, which organizes the gauge identities and the physical cohomology; its quantum consistency further requires preservation of BRST symmetry by the measure and regulator.

To define the BRST transformations, replace the infinitesimal bosonic gauge parameter  by the Grassmann-odd ghost coefficient. We use a single convention, $\delta_{\rm BRST}=\epsilon s$, with constant Grassmann parameter $\epsilon$ and nilpotent differential $s$. For nonlinear target coordinates the fundamental transformation is

\begin{equation}
\delta_{\rm BRST}\pi^a=g\epsilon\,c^I(x)\zeta_I^a(\pi).
\label{eq:brst_goldstone_killing}
\end{equation}

A form proportional to $T_I\pi$ is recovered only where the action on the scalar variables is linear.

Nilpotency must be imposed on the differential $s$, not on $\delta_{\rm BRST}=\epsilon s$ itself, since a single Grassmann parameter satisfies $\epsilon^2=0$. We therefore define

\begin{equation}\label{eq1}
\begin{aligned}
sA_\mu^I&=(D_\mu c)^I, & s\pi^a&=g c^I\zeta_I^a,\\
sc^I&=-\frac g2 f_{JK}{}^I c^Jc^K, & s\bar c_I&=B_I,\qquad sB^I=0.
\end{aligned}
\end{equation}
The infinitesimal transformation is then $\delta_{\rm BRST}\Phi=\epsilon\,s\Phi$. $a,b,\ldots$ denote target-space indices, whereas $I,J,K,\ldots$ denote gauge-algebra indices. Now the nilpotency is

\begin{equation}
s^2A_\mu^I=s^2c^I=s^2\bar c_I=s^2B^I=s^2\pi^a=0,
\end{equation}
which is off shell: the $B^I$ field removes any need to use its equation of motion. Algebraically, $s^2c^I=0$ and $s^2A_\mu^I=0$ follow from the Jacobi identity, while $s^2\pi^a=0$ follows from the Killing-vector closure. For the Goldstone sector, BRST nilpotency follows from the compatibility between the ghost transformation and the Lie-algebra representation realized by the gauged Killing vectors. The detailed derivation is given below in the Killing interpretation of the BRST charge.

The BRST transformations defined above ensure the invariance of the Lagrangian of Eq.~\eqref{lagrangiano_BRST_aux}  under these transformations, thus replacing the original local gauge symmetry, which is explicitly broken after gauge fixing, by a global fermionic symmetry known as BRST symmetry~\cite{becchi1976renormalization}.

The BRST transformation of the GBs can be formalized through the action of the Killing vectors $\zeta^a_b(\pi)$, originally obtained in Eq.~\eqref{eq:killing_inf}, which generate the isometries of the scalar field space. The transformation for the GBs is
$\delta_{BRST} \pi^a = g \epsilon c^b(x) \zeta_b^a(\pi)$, 
where $\zeta_b^a(\pi)$ describes the infinitesimal generators of the internal symmetry. This relation shows  the deep connection between the geometric structure of the configuration space and the algebraic formulation of BRST transformations~\cite{Henneaux:1998hq,BaulieuThierryMieg1982}.

The BRST procedure, in this particular model, arises as a fundamental mechanism for consistently treat the gauge redundancies during quantization, especially after the introduction of the gauge-fixing term and ghost fields through the FP method~\cite{tyutin1975gauge, becchi1976renormalization}. Although local gauge invariance is explicitly broken, the BRST —nilpotent and global symmetries are respecting the mathematical and physical integrity of the theory.

\subsection{BRST Invariance}

In this section, we show explicitly as BRST invariance is manifest in all terms of Lagrangian from Eq. \ref{lagrangiano_BRST_aux}. 
A compact way to establish BRST invariance is to organize the gauge-fixing and ghost sector as a BRST-exact term. To distinguish the local density from the integrated gauge-fixing fermion, let us define
\begin{equation}
\psi_{\rm gf}=\bar c_I\left(F^I(A,\pi)+\frac{\xi}{2}B^I\right),\qquad \Psi=\int d^4x\,\psi_{\rm gf}.
\end{equation}

Using $s\bar c_I=B_I$, $sB_I=0$, one obtains for the gauge fixing and ghost terms 
\begin{equation}
\mathcal L_{\rm gf+gh}=s\psi=B_I F^I+\frac{\xi}{2}B_IB^I-\bar c_I\,sF^I.
\end{equation}
For $F^I=\partial\!\cdot\!A^I$, this reduces to the Lorenz-gauge expression used for the explicit current of the following section. Because $s^2=0$ off shell in the presence of $B^I$, the pplication of $s$ operator

\begin{equation}
s\mathcal L_{\rm gf+gh}=s^2\Psi=0.
\end{equation}

The Yang-Mills term is BRST invariant because $sF_{\mu\nu}^I=g f^{I}_{JK}F_{\mu\nu}^J c^K$, while the Goldstone kinetic term is invariant because the gauged vector fields are Killing vectors of the target-space metric. Hence for the Lagrangian of Eq. \eqref{eq:lagrangiano BRST con bosones de goldstone}, we obtain the following

\begin{equation}
s\mathcal L_{ext}=0.
\end{equation}

Both $f^{I}_{JK}$ and $c^J c^K$ are antisymmetric in $J,K$, so their contraction is generally nonzero and is precisely required by the non-Abelian BRST algebra.

\subsection{Physical Interpretation and Role of BRST Symmetry}

As we studied above, the BRST symmetry is a fundamental global fermionic symmetry that naturally emerges when quantizing gauge theories after gauge fixing and the introduction of ghost fields via the FP method~\cite{tyutin1975gauge,becchi1976renormalization}. Unlike the original local gauge symmetry, the BRST symmetry is global (the parameter $\epsilon$ is constant and anticommuting), and its transformation acts jointly on both physical fields and ghost fields~\cite{henneaux1992quantization,PiguetSorella1995}.

The crucial importance of BRST symmetry lies in the fact that it allows for the definition of the BRST charge, a conserved and nilpotent quantity (i.e. $Q^{2}_{\text{BRST}}=0$)~\cite{kugo1979local}, which enables the construction of the physical state space via the cohomology of this charge. This means that physical states are those that are invariant under the BRST transformation but are not trivially generated by it (i.e., they are not BRST-exact states)~\cite{ kugo1979local,henneaux1992quantization}. In this way, gauge-redundant degrees of freedom are removed cohomologically. Positivity and unitarity of the physical sector require the standard additional assumptions on the BRST charge and quartet mechanism, while a four-dimensional NL$\sigma$M remains an effective field theory rather than a power-counting-renormalizable model.

Therefore, BRST invariance is a cornerstone of covariant gauge quantization on three aspects: it organizes the gauge identities, permits a cohomological definition of the physical sector, and constrains admissible quantum corrections~\cite{PiguetSorella1995,BarnichBrandtHenneaux2000}. At the quantum level of the theory, conservation and nilpotency require an anomaly-free BRST symmetry. In the standard antifield formulation of the nonlinear BRST transformations, quantum BRST invariance is described in the Slavnov-Taylor functional identity:

\begin{equation}
\mathcal S(\Gamma)=0\quad\text{(anomaly free)},\qquad
\mathcal S(\Gamma)=\hbar\Delta^{(1)}+\mathcal O(\hbar^2)\ \Rightarrow\ \mathcal B_{S_0}\Delta^{(1)}=0.
\end{equation}
The two expressions stablish the following interpretations: the second parametrizes a possible one-loop breaking whose nontrivial cohomology would obstruct the first. Here $\mathcal B_{S_0}$ is the linearized Slavnov operator; a nontrivial ghost-number-one $\Delta^{(1)}$ signals a BRST anomaly. This fact is stated only as the general BRST consistency condition; no perturbative gauge anomaly is identified for the purely bosonic field content of gauged NL$\sigma$M. Consequently, BRST symmetry provides the formal framework to separate the physical degrees of freedom from the artifacts introduced during gauge fixing.

For the explicit Noether-current calculation we now specialize to the Lorenz choice $F^I=\partial\!\cdot\!A^I$; the later cohomological statements do not depend on this particular gauge-fixing choice.

\section{Construction of the BRST operator}

The BRST formalism is based on the existence of a residual global fermionic symmetry that remains after the gauge-fixing procedure, in which ghost fields are introduced~\cite{Nemeschansky:1988}. This symmetry is characterized by an infinitesimal anticommuting operator, known as the BRST operator, whose action on the fields defines the BRST transformation. This operator can be explicitly constructed by applying Noether's theorem to residual the global fermionic symmetry~\cite{henneaux1992quantization}.

Let us recall the extended Lagrangian in Eq.~\eqref{lagrangiano_BRST_aux}, where a warning is required in applying the standard first-order Noether formula because the ghost density $-\bar c_I\partial^\mu(D_\mu c)^I$ contains second derivatives of $c^a$ if it is used literally. Up to a total derivative one may instead use
\begin{equation}
-\bar c_I\partial^\mu(D_\mu c)^I\simeq (\partial^\mu\bar c_I)(D_\mu c)^I.
\end{equation}
For this first-order representative the BRST variation differs from zero by a total derivative, $s\mathcal L'_{ext}=\partial_\mu K^\mu$, with
\begin{equation}
K^\mu=B_I(D^\mu c)^I.
\end{equation}
The Noether current is therefore
\begin{equation}
j^\mu_{\rm BRST}=\sum_{\Phi}\frac{\partial\mathcal L'_{ext}}{\partial(\partial_\mu\Phi)}s\Phi-K^\mu,
\end{equation}
where left/right Grassmann derivatives are understood consistently. The antighost contribution $B_I(D^\mu c)^I$ cancels the improvement term $K^\mu$, and one convenient representation of the BRST-current is
\begin{equation}\label{eq:current}
j^\mu_{\rm BRST}
=g\,g_{ab}(\pi)D^\mu\pi^b\,c^I\zeta_I^a
+\left(-F_I^{\mu\nu}+B_I\eta^{\mu\nu}\right)(D_\nu c)^I
-\frac g2(\partial^\mu\bar c_I)f^I{}_{JK}c^Jc^K.
\end{equation}
On the equations of motion, $\partial_\mu j^\mu_{\rm BRST}=0$. The current is defined up to conserved improvement terms and terms proportional to the field equations.
With vanishing BRST-current flux at spatial infinity, $dQ_{\rm BRST}/dt=0$; boundary contributions must be retained for \emph{nontrivial representations of them}.

The current form of Eq.~\ref{eq:current},  obtained via Noether's theorem, is fundamental for defining the BRST charge that ensures the consistency and invariance of the gauge theory under a quantization procedure. Finally, in this aspect, the BRST current organizes the symmetry constraints of the gauge-fixed theory. It does not by itself prove unitarity or renormalizability; those require the cohomological and EFT conditions stated below~\cite{kugo1979local,Brandt:1996mh}.

\subsection{BRST Charge}

The BRST charge is defined from the BRST current—whose explicit form was calculated previously—as the spatial integral of its temporal component, i.e.,
under the standard equal-time graded brackets it generates the symmetry, $s\Phi=i[Q_{\rm BRST},\Phi\}_{\rm gr}$, up to the overall sign convention for $Q_{\rm BRST}$ \cite{kugo1979local},

\begin{equation}
Q_{\text{BRST}} = \int d^3x\, j_{\text{BRST}}^0(x).
\end{equation}

Substituting the previously obtained expression for the BRST current, 
we obtain the explicit expression for the BRST charge:

\begin{equation}\label{eq:QBRST_final}
Q_{\text{BRST}} = \int d^3x \left[ g g_{ab}(\pi)\zeta_I^a c^I D^0\pi^b
+\left(-F_I^{0\nu}+\eta^{0\nu}B_I\right)(D_\nu c)^I
-\frac{g}{2}(\partial^0\bar c_I)f^I{}_{JK}c^Jc^K \right].
\end{equation}
With the generator relation above, off-shell $s^2=0$ implies $Q_{\rm BRST}^2=0$ modulo possible anomalous or boundary obstructions.

The first term in Eq.~\eqref{eq:QBRST_final} constructs a direct connection between the BRST charge and the geometry of the target space. The target space formed the metric $g_{ab}(\pi)$, the Killing vectors $\zeta_i^a(\pi)$, and the covariant time derivative of the Goldstone fields. From a geometrical description of the BRST procedure, this term represents the projection of the scalar-field dynamics onto the directions tangent to the symmetry orbits of the target manifold defined by the Killing flow. The presence of the ghost field $c^I$ indicates that this pairing does not correspond to an ordinary gauge transformation, but rather to its graded extension in field space. This implies that the ghost field $c^{I}$
replaces the ordinary gauge parameter, converting the gauge transformation into a Grassmann-graded transformation along the same gauge-orbit directions. In this approach, the BRST charge acts as the generator of the Grassmann-odd differential along gauge-orbit directions.

The second term contains the contribution of the gauge sector and the gauge-fixing procedure. The combination of the field strength tensor $F_{\mu\nu}^I$, the auxiliary field $B_I$, and the covariant derivative of the ghost field $D_\nu c^I$ shows that BRST symmetry is not constrained to the scalar sector uniquely, but extends to the full gauge structure of the theory. From a physical point of view, this term reflects how the gauge redundancy of the classical system is reorganized, after quantization, into a global fermionic symmetry of the extended field space. In this way, the BRST charge also involves the dynamics of the additional degrees of freedom required for a consistent quantization of the theory.

The third term explicitly contains the internal algebraic structure of the gauge group through the structure constants $f^{I}_{ JK}$, whose presence indicates that BRST transformations do not act as independent displacements along each symmetry direction, but rather as operations whose composition is governed by the underlying Lie algebra. Therefore, the BRST charge depends not only on the differential geometry of the target space, but also on the algebraic structure of the gauge and ghost sectors. Altogether, $Q_{\mathrm{BRST}}$ synthesizes in a single object the geometry of the symmetry orbits, the gauge dynamics, and the algebraic-cohomological organization of the theory.

Although the expression of Eq.~\eqref{eq:QBRST_final} cannot generally be integrated analytically due to the geometric and nonlinear complexity of the theory, it enables the study of crucial algebraic and structural properties such as nilpotency, conservation quantities, and the cohomological classification of physical states.

\subsection{Killing interpretation of the BRST charge}

In the BRST construction, the bosonic gauge parameters $\alpha^I$ are replaced by Grassmann-odd ghost fields $c^I(x)$ in the differential $s$; the separate constant Grassmann parameter $\epsilon$ only forms $\delta_{\rm BRST}=\epsilon s$

\begin{equation}
s\,\pi^a = g\,c^I\zeta_I^a(\pi),
\label{eq:brst_pi}
\end{equation}

where $s$ is the BRST differential. In this formulation, the ghost fields are Grassmann-valued coefficients along the gauge directions, while $s$ acts as a differential on the extended field space \cite{Henneaux:1998hq}.

The nilpotency check  starts with

\begin{equation}
s^2\pi^a=g(sc^I)\zeta_I^a-gc^I s(\zeta_I^a),\qquad
s(\zeta_I^a)=g c^J\zeta_J^b\partial_b\zeta_I^a .
\end{equation}

using the operation over ghost fields

\begin{equation}
sc^I=-\frac g2 f_{JK}{}^I c^Jc^K,
\label{eq:brst_ghost_b}
\end{equation}

one finds

\begin{equation}
s^2\pi^a=-\frac{g^2}{2}c^Ic^J
\left(f_{IJ}{}^K\zeta_K^a-[\zeta_I,\zeta_J]^a\right)=0 .
\end{equation}
Thus BRST nilpotency on the Goldstone sector is the graded realization of Killing-vector closure. Because $\zeta_i^a$ depends on the target coordinates $\pi^a$, the action of $s$ on the Killing fields gives
\begin{equation}
s(\zeta_i^a)=s(\pi^b)\partial_b\zeta_i^a
=
g\,c^j\zeta_j^b\partial_b\zeta_i^a.
\end{equation}

On the other hand, the BRST transformation of the fields is given by 
\begin{equation}
s\,c^i=-\frac{g}{2}f_{jk}{}^{i}c^j c^k.
\label{eq:brst_ghost}
\end{equation}

By substituting these expressions, we can get
\begin{equation}
s^2\pi^a
=
-\frac{g^2}{2}f_{jk}{}^{i}c^j c^k\zeta_i^a
-
g^2 c^i c^j \zeta_j^b\partial_b\zeta_i^a.
\end{equation}

Using the anticommuting property of the ghost fields, the second term can be written as

\begin{equation}
g^2 c^i c^j \zeta_j^b\partial_b\zeta_i^a
=
\frac{g^2}{2}c^i c^j
\left(
\zeta_j^b\partial_b\zeta_i^a-\zeta_i^b\partial_b\zeta_j^a
\right),
\end{equation}

and recognizing the Lie bracket among Killing vectors, we can obtain 

\begin{equation}
\label{eq:mainKF}
s^2\pi^a=-\frac{g^2}{2}c^Ic^J
\left(f_{IJ}{}^K\zeta_K^a-[\zeta_I,\zeta_J]^a\right)=0 .
\end{equation}
Thus BRST nilpotency on the Goldstone sector is the graded realization of Killing-vector closure.

Thus BRST nilpotency on the Goldstone sector is the graded realization of Killing-vector closure. The result allows a geometrical interpretation: nilpotency of the BRST differential on the GBs encodes the fact that successive graded gauge displacements along the Killing directions do not generate a new independent direction transverse to the gauge orbit, but close within the same Killing generated orbit. Therefore, Killing-vector closure provides the geometric input required for BRST nilpotency in the scalar sector, which in turn makes the cohomological construction possible..

In this way, the BRST charge acquires a clear geometrical status: it is the graded generator of the infinitesimal gauge flow on the extended field space, simultaneously encoding the isometries of the scalar manifold, the Lie-algebraic structure of the gauge sector, and the cohomological organization of the physical Hilbert space, as we will below.

\subsubsection{Illustrative example on $O(4)/O(3)\simeq S^3$}
One particular example for the relation between nilpotency and Killing fields closure property is the model with the coset $O(4)/O(3)\simeq S^3$. Here the unbroken {$SO(3)$} group acts on the three Goldstone coordinates through the Killing fields {$\zeta_i^a=\epsilon_i{}^{ab}\pi_b$}, so that {$[\zeta_i,\zeta_j]^a=\epsilon_{ij}{}^k\zeta_k^a$}. Consequently, {$s\pi^a=g c^i\zeta_i^a$} together with {$s c^i=-\frac{g}{2}\epsilon^i{}_{jk}c^jc^k$} gives {$s^2\pi^a=0$}, making the relation between Lie-algebra closure and BRST nilpotency, Eq.\eqref{eq:mainKF},  completely explicit for these $SO(3)$ symmetry orbits.

\subsection{Physical Interpretation of the BRST Charge}

The BRST charge, defined as the spatial integral of the temporal component of the BRST current according to Eq.~\eqref{eq:QBRST_final}, carries a deep and fundamental physical meaning in quantized gauge theories. Below, we outline this interpretation through the following key points:

\begin{itemize}
    \item Generator of the global BRST symmetry:
    The BRST charge generates the global BRST symmetry on the extended state space. Its nilpotent action organizes this space into BRST-closed and BRST-exact sectors, while the physical state space is identified with the ghost-number-zero cohomology $H_{0}(Q_{\text{BRST}})$.

    \item Selection of physical states:
    The BRST charge clearly distinguishes between physical and unphysical states in the theory. At ghost number zero, physical representatives are BRST closed,

    \begin{equation}
        Q_{\text{BRST}}|\text{physical}\rangle=0.
    \end{equation}

    with the additional identification of representatives that differ by a BRST-exact state. Thus the physical state space is a cohomology rather than the full kernel of $Q_{\rm BRST}$.

    \item Nilpotency and quantum consistency:
    The BRST charge satisfies the fundamental property of nilpotency,

    \begin{equation}
        Q_{\text{BRST}}^2=0.
    \end{equation}

   Nilpotency provides the algebraic foundation for the BRST cohomology. Unitarity and positivity of the physical sector additionally require the appropriate adjoint structure, boundary conditions, quartet mechanism, and absence of a BRST anomaly.

    \item Elimination of redundant degrees of freedom:
    The BRST charge emerges precisely as a mechanism to handle gauge redundancies. Physically, BRST-closed states are admissible representatives, while BRST-exact states represent trivial cohomology classes associated with gauge redundancy.

    \item Quantum-level consistency of the theory:
    Finally, an anomaly-free BRST symmetry constrains the quantum theory through its gauge identities and cohomology. In the NL$\sigma$M this constrains counterterms order by order in the effective expansion rather than rendering the theory power-counting renormalizable. This makes the BRST charge an essential element in the modern mathematical and physical formulation of gauge theories, such as quantum chromodynamics (QCD) and the electroweak theory.

    \item Commuting property of $Q$: Another fundamental property of the BRST charge is that it commutes with the Hamiltonian of the theory:

\begin{equation}
[Q,H]=0,
\end{equation}
\end{itemize}

All statements about $Q_{\rm BRST}^2=0$, conservation of $Q_{\rm BRST}$, and $[Q,H]=0$ at the quantum level are understood to require a BRST-preserving regulator and measure, absence of anomalies, and suitable boundary conditions.

From the physical point of view, the nilpotency of the BRST charge,
\begin{equation}
Q_{\mathrm{BRST}}^2=0,
\end{equation}
is the fundamental property that guarantees the consistent elimination of gauge redundancies at the quantum level. Nilpotency implies that the BRST transformation does not generate an infinite chain of independent gauge descendants, but instead closes algebraically after one nontrivial action. As a consequence, physical states are defined by the condition
\begin{equation}
Q_{\mathrm{BRST}}\,|\mathrm{phys}\rangle =0,
\end{equation}
whereas states of the form
\begin{equation}
|\psi\rangle = Q_{\mathrm{BRST}}|\chi\rangle
\end{equation}
represent the trivial cohomology class and therefore do not define independent physical states. In this way, the physical Hilbert space is naturally identified with the cohomology of the BRST charge.

Geometrically, this structure reflects the fact that the Killing vectors generate the gauge orbits in field space, while the BRST charge implements, at the quantum level, the reduction along those symmetry directions. The $Q_{\mathrm{BRST}}$ nilpotency expresses that the graded gauge flow generated on the extended field space does not produce new physical directions transverse to the gauge orbit, but remains entirely confined within the same redundancy structure. In this sense, BRST cohomology provides the quantum realization of the quotient between field configurations related by the action of the Killing directions.

\begin{figure}
\centering
\includegraphics[scale=0.16]{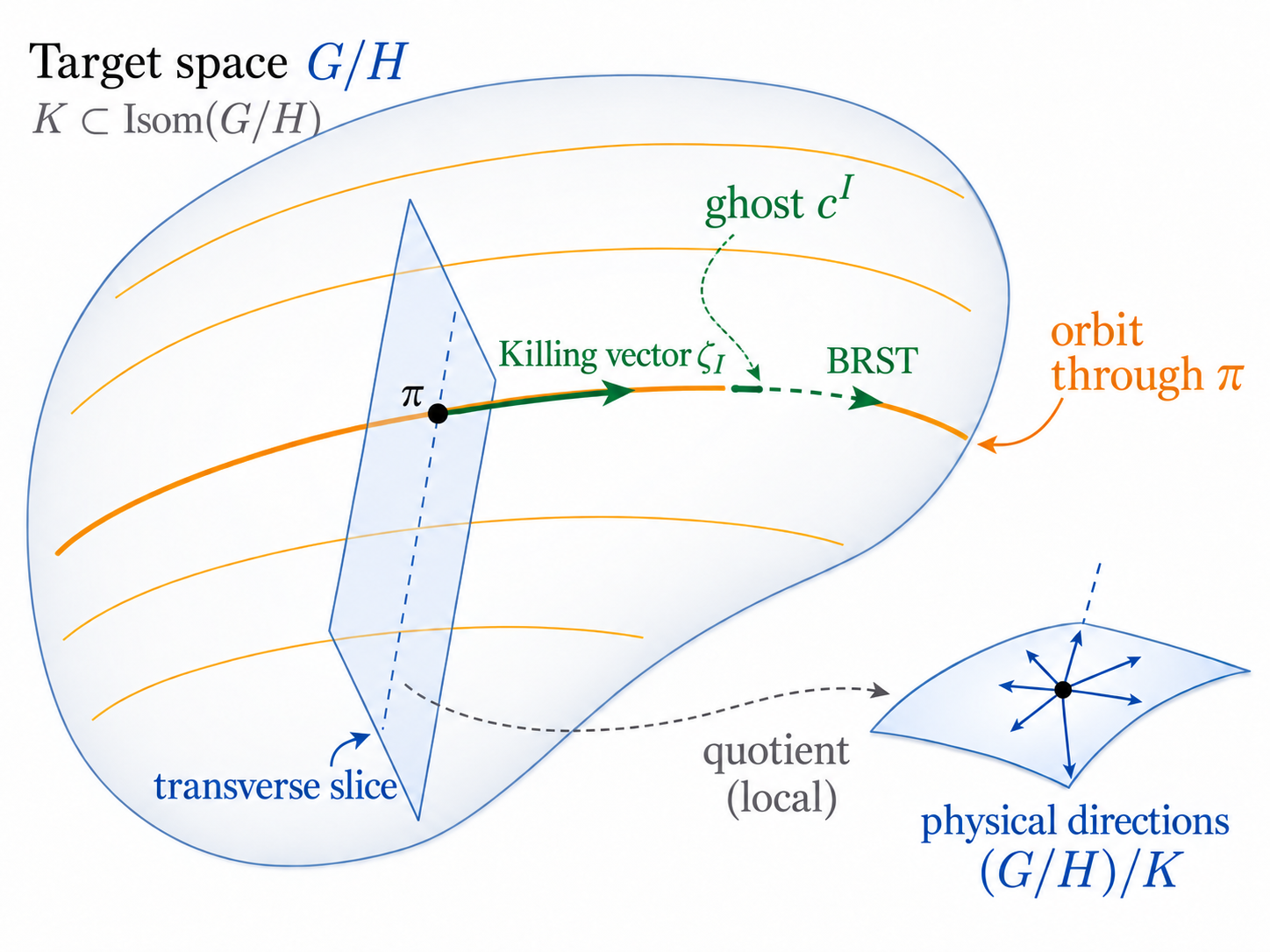}
\caption{Geometric interpretation of Killing directions and BRST reduction for the NL$\sigma$M and spaces generated after SSB. The Killing vectors ($\zeta_I$) yields the infinitesimal action of the symmetry group $K$ on the target space ($G/H$) and are tangent to the symmetry orbits. After gauge fixing procedure, the BRST transformation follows these same symmetry directions, with the ghost fields ($c^I$) providing their Grassmann-odd extension. A local transverse plane on $\pi$ represents the separation between gauge directions and locally independent configurations while the reduced space illustrates the geometrical counterpart of eliminating the gauge redundancies with the BRST-cohomology with the BRST-cohomology. The graphical representation for this geometric interpretation was developed with the assistance of generative AI tools (GPT 5.6-model).}
\label{fig:BRST-Killing}
\end{figure}

In summary, the Killing vectors describe the geometrical directions of symmetry of the target space; their commutation relations determine how these directions close and compose; the BRST operator explores them in a graded way through the ghost sector \cite{Henneaux:1998hq, henneaux1992quantization, BaulieuThierryMieg1982}. And finally, the BRST charge generates this flow globally on the extended field space. The nilpotency of the BRST charge may therefore be understood as the cohomological manifestation of the geometrical closure of the isometries together with the algebraic consistency of the gauge and ghost sectors, thereby providing a criterion for the identification of the physical states of the theory. The effect of Killing vectors and its influence to the BRST-cohomology can be seen in the Fig. \ref{fig:BRST-Killing}

Within a NL$\sigma$M, the BRST charge is interpreted as a structural operator describing the gauge symmetry on a curved internal field space. Since the scalar fields parametrize a nonlinear target manifold endowed with a metric and Killing vectors defining the gauge orbits, the BRST charge implements gauge transformations as nilpotent fermionic generators acting along the corresponding symmetry directions. Indeed, the nilpotency reflects the closure of the gauge algebra and the geometric consistency of the target-space isometries defined by the organization of the Killing vectors. Besides, the cohomological action provides a criterion for identifying physical states by eliminating gauge redundancies at the quantum level. In this way, the BRST charge encapsulates the formal relation between gauge symmetry, internal geometry, and the structure of the physical Hilbert space, as we will see below \cite{FischHenneauxStasheffTeitelboim1989}.

In summary, the BRST charge is not merely a formal object but a physical object that describes an internal consistency and the correct identification of physical states. It is key to the current understanding of quantized non-Abelian gauge theories. Therefore, the BRST charge, although given as a spatial integral, encodes all the physical and algebraic structure that ensures the proper quantization of the gauge theory~\cite{kugo1979local,
henneaux1992quantization,hull1990brst}.

\section{Structure of the Physical State Space from the BRST Operator}

The conserved nilpotent BRST charge organizes the extended state space by ghost number. In this section we set $Q\equiv Q_{\rm BRST}$ to avoid switching between two symbols for the same operator. Let $\mathcal{H}_{ext}$ denotes the BRST-extended state space containing both physical and gauge-dependent sectors, including the ghost and auxiliary degrees of freedom
\begin{equation}
\mathcal H_{\rm ext}=\bigoplus_{n\in\mathbb Z}\mathcal H_n,
\qquad
Q:\mathcal H_n\longrightarrow\mathcal H_{n+1},
\qquad Q^2=0.
\end{equation}
Here, $\mathcal{H}_{n}$ denotes the subspace of states with ghost number $n$, so that the extended state space is graded by the integer-valued ghost number. Define the closed and exact subspaces
\begin{equation}
Z^n(Q)=\ker\!\left(Q:\mathcal H_n\to\mathcal H_{n+1}\right),
\qquad
B^n(Q)=\operatorname{im}\!\left(Q:\mathcal H_{n-1}\to\mathcal H_n\right).
\end{equation}
Nilpotency implies $B^n(Q)\subseteq Z^n(Q)$, and the BRST cohomology is
\begin{equation}
H^n(Q)=\frac{Z^n(Q)}{B^n(Q)}.
\end{equation}
The physical state space is the ghost-number-zero cohomology,
\begin{equation}\label{eq:Hphys_compact}
{\mathcal H_{\rm phys}=H^0(Q_{\rm BRST})
=\frac{\ker(Q:\mathcal H_0\to\mathcal H_1)}{\operatorname{im}(Q:\mathcal H_{-1}\to\mathcal H_0)}}.
\end{equation}
Thus a physical state is represented by a BRST-closed vector $|\psi\rangle\in\mathcal H_0$, with the equivalence relation
\begin{equation}
|\psi\rangle\sim|\psi\rangle+Q|\chi\rangle,
\qquad |\chi\rangle\in\mathcal H_{-1}.
\end{equation}
The physical space is the quotient cohomology $H^0(Q_{\rm BRST})$, not merely $\ker Q_{\rm BRST}$.

The metric properties require an additional assumption beyond nilpotency. If $Q$ is self-adjoint with respect to the appropriate indefinite/Krein inner product, then for a closed state $Q|\psi\rangle=0$ and an exact state $|\phi\rangle=Q|\chi\rangle$,
\begin{equation}
\langle\psi|\phi\rangle=\langle Q\psi|\chi\rangle=0,
\end{equation}
and two exact states satisfy
\begin{equation}
\langle Q\chi_1|Q\chi_2\rangle
=\langle\chi_1|Q^2|\chi_2\rangle=0.
\end{equation}
Accordingly, the nullness and decoupling of exact states follow from nilpotency together with the appropriate adjoint structure; positivity of the final physical Hilbert space further relies on the standard BRST assumptions~\cite{kugo1979local,
henneaux1992quantization,Nemeschansky:1988}.

Physical observables admit the analogous cohomological characterization. A ghost-number-zero operator $\mathcal O$ is BRST closed when
\begin{equation}
[Q,\mathcal O\}_{\rm gr}=0,
\end{equation}
and operators differing by a BRST-exact term act identically on cohomology,
\begin{equation}
\mathcal O\sim \mathcal O+[Q,X\}_{\rm gr}.
\end{equation}
This relation explains, under a BRST-invariant measure and suitable boundary conditions, why BRST-exact changes of the gauge-fixing prescription do not modify physical matrix elements~\cite{BarnichBrandtHenneaux2000}. At any fixed EFT order, admissible local counterterms lie in ghost-number-zero BRST cohomology, modulo BRST-exact terms, equations of motion, and total derivatives~\cite{PiguetSorella1995,
BarnichBrandtHenneaux2000,JoglekarLee1976}.
Ghost fields are therefore best viewed as Grassmann-valued variables that encode the gauge-orbit directions in the extended field space. Their next to leading contributions participate in the cancellation of gauge-dependent unphysical modes and in the Slavnov-Taylor identities, while the physical content is obtained only after passing to $H^0(Q_{\rm BRST})$~\cite{LeeZinnJustin1972I, ZinnJustin1975}. At the quantum level this construction presupposes that BRST symmetry is non-anomalous; otherwise conservation or nilpotency of the quantum charge can be obstructed.

From a geometrical perspective, it is possible to describe the Killing vectors span the local gauge-orbit directions, the ghosts supply Grassmann coefficients along those directions~\cite{BaulieuThierryMieg1982}, and $Q_{\rm BRST}$ generates the corresponding graded flow. The quotient in Eq.~\eqref{eq:Hphys_compact} is therefore the quantum-cohomological counterpart of removing gauge-orbit redundancy, while retaining the physical equivalence classes.

\section{Conclusions}

We have presented a detailed formulation of the BRST formalism applied to non-Abelian gauge theories that include Goldstone bosons. The explicit derivation of the BRST current from the extended Lagrangian of a NL$\sigma$M and its detailed analysis have made it possible to study the algebraic and geometric foundations that underlie the consistency of these quantum theories.

It has been shown that the BRST charge for the NL$\sigma$M, constructed through the integration of the conserved current, acts as the generator of a global fermionic symmetry whose nilpotency implies a cohomological structure in the space of states. This feature is a critical to define formally the physical subspace of the Hilbert space, leading to separating in an unambiguous way the physical degrees of freedom from the redundant ones associated with gauge symmetry.

Likewise, the geometry of the coset space in the NL$\sigma$M, reflected in the interaction between GBs and gauge fields through Killing vectors and internal metrics, is naturally involved within the BRST formalism, describing a clear and consistent interpretation of gauge orbits in the presence of spontaneously broken symmetries. 

In the geometrical approach presented here, the closure of the Killing-vector algebra provides the target-space realization of BRST nilpotency in the Goldstone sector. It implies that successive graded displacements along the Killing directions close within the same gauge orbit rather than generating a new transverse physical direction. The full nilpotency of the BRST charge additionally depends on the gauge and ghost algebras, the Jacobi identity, and, at the quantum level, the absence of a BRST anomaly.

Finally, the BRST forms a robust mathematical framework to interpret gauge theories with scalar degrees of freedom. When the quantum BRST symmetry is anomaly free and the standard conditions for the physical cohomology are satisfied, it is converted into a consistent treatment of unitarity and gauge-independent observables. For a four-dimensional NL$\sigma$M, these statements must be seen within its effective-field-theory domain: BRST symmetry constrains the allowed counter-terms from the expansion in momenta $p/\Lambda_{\rm EFT}$ but does not by itself make the theory power-counting renormalizable.

In summary, the BRST formalism provides the conceptual and technical framework for a consistent perturbative BRST formulation of gauge theories with Goldstone modes. In four-dimensional NL$\sigma$Ms this statement is restricted to the EFT domain and does not address global Gribov obstructions.

\subsection*{Acknowledgments}
Andrés Castillo acknowledges the financial support from the internal grants of Vicerrectoria de Ciencia, Tecnología e Innovación (Universidad Antonio Nari\~no) for the projects "\emph{Efectos de cosmologías no estándar en la detección del fondo difuso de neutrinos de supernovas (DSNB)}" 2025-205 and "\emph{Fenomenología de campos vectoriales en regímenes extremos: De estrellas de neutrones a escalas cosmológicas}" 2026-223. Generative artificial intelligence tools provided through OpenAI-GPT (5.6-sol model) were used to assist the diagrammatic design of the main figure, as well as for grammatical corrections for the text.

\bibliographystyle{ieeetr}
\bibliography{references}
\end{document}